\documentclass{jps-cp}
\usepackage{graphicx}
\usepackage{bm,color}
\usepackage{braket}
\usepackage{amsmath,amssymb}

\title{Towards Reliable Nuclear Matrix Elements
	for Neutrinoless $\beta\beta$ Decay}

\author{Javier \textsc{Men{\'e}ndez}}

\inst{Center for Nuclear Study, The University of Tokyo, Tokyo 113-0033, Japan}

\email{menendez@cns.s.u-tokyo.ac.jp}

\recdate{March 27, 2018}

\abst{The calculated nuclear matrix elements for the neutrinoless double-beta ($0\nu\beta\beta$) decay suffer from several limitations. Predicted matrix-element values depend on the many-body method used to calculate them and, in addition, they may need to be ``quenched", albeit by an unknown amount due to the relatively high momemtum transferred in the decay. These uncertainties are hard to figure out from theoretical calculations alone because of the unique character of $0\nu\beta\beta$ decay, not clearly related to other nuclear structure observables. We present recent progress in the determination of the nuclear matrix elements. First, we report improved shell model calculations in an extended configuration space of two harmonic oscillator shells. Second, we note the relation between the $0\nu\beta\beta$ decay and double Gamow-Teller transitions that can in principle be measured in double charge-exchange reactions. These new insights pave the way for a more reliable estimation of the value of the nuclear matrix elements.}

\kword{double-$\beta$ decay, nuclear matrix element, shell model, charge-exchange reactions}

\begin{document}
\maketitle

\section{Introduction}
In a neutrinoless double-beta ($0\nu\beta\beta$) decay an atomic nucleus decays into another one with two more protons and two fewer neutrons, emitting two electrons. In other words, two leptons are created. Such violation of the lepton number conservation is only possible if neutrinos---unlike any other fundamental particle---are its own antiparticle, a possibility first suggested by Ettore Majorana in the 1930's. In spite of the challenges associated with the detection of a process that involves new physics, $0\nu\beta\beta$ decay is being pursued by several experimental collaborations~\cite{KamLAND-Zen16,EXO18,GERDA18,MAJORANA18,CUORE18,NEXT18,CUPID18}. A major advantage is that the parameter $m_{\beta\beta}$ that controls the $0\nu\beta\beta$ decay half-life is fully fixed by the the known neutrino mass differences and mixing angles, in such a way that $m_{\beta\beta}$ only depends on the ordering ---``normal" or ``inverted"---of the neutrinos masses~\cite{engelmen_review}:
\begin{equation}
\label{eq:half-life}
[T^{0\nu}_{1/2}]^{-1}=G^{0\nu}
\left|M^{0\nu\beta\beta}\right|^2  m_{\beta\beta}^2\,.
\end{equation}
There is, however, a catch. The $0\nu\beta\beta$ decay half-life also depends on the value of an associated nuclear matrix element (NME), $M^{0\nu\beta\beta}$, like any other {\it nuclear} decay---$G^{0\nu}$ is a known phase-space factor. NMEs need to be calculated theoretically, and their value is key to assess the prospects to observe $0\nu\beta\beta$ decay in present and next-generation experiments.

At present, predicted NME values vary by a factor two or three depending on the many-body method used to calculate them. In addition, the results may need to be ``quenched" as is common for $\beta$ decays, but since the momentum transfer in $0\nu\beta\beta$ decay is much larger, the necessity of such ``quenching" is unclear~\cite{engelmen_review}. These proceedings discuss recent ideas towards a more reliable determination of the 
NMEs, with focus on improved many-body calculations, and on the relation between $0\nu\beta\beta$ decay and double Gamow-Teller (GT) transitions.

\section{Shell model nuclear matrix elements in two harmonic oscillator shells}

Among the nuclear many-body methods used to study $0\nu\beta\beta$ decay, the nuclear shell model plays a prominent role, as one of the most successful approaches to nuclear structure~\cite{cau05}. Nonetheless the main drawback of shell-model NMEs is that they are typically calculated limiting the configuration space to one harmonic oscillator shell. While, in general, such restriction works very well to describe the nuclear structure and spectroscopy of stable nuclei, it has been claimed that such a configuration space may not be large enough to obtain converged $0\nu\beta\beta$ decay NMEs~\cite{vogel12}.

The lightest $\beta\beta$ emitter is $^{48}$Ca. This is therefore the nucleus for which shell model calculations beyond one harmonic oscillator shell are  less demanding computationally. Ref.~\cite{iwata16} calculated the NME for the $0\nu\beta\beta$ decay in a configuration space consisting of two harmonic oscillator shells, the $sd$ and $pf$ shells. Previous shell model calculations were restricted to the $pf$ shell, while Ref.~\cite{iwata16} was able to include up to $2\hbar\omega$ $sd$-$pf$ excitations. The calculation was validated by reproducing the excitation spectra of the initial and final nuclei of the decay, $^{48}$Ca and $^{48}$Ti~\cite{iwata16}. In addition, the shell model calculation of Ref.~\cite{iwata16} showed a good description of the GT strength, including the GT giant resonance (GR), of $^{48}$Ca and $^{48}$Ti into $^{48}$Sc~\cite{iwata15}---these GT strengths had been measured in charge-exchange experiments~\cite{yako-ca48}---, and reproduced the two-neutrino $\beta\beta$ decay matrix element of $^{48}$Ca as well. For the transition operators, the agreement to experiment was only possible after a ``renormalization", or ``quenching", of the theoretical predictions by a factor $q=0.71$ for each spin-isospin ${\bm \sigma}\tau$ term present in the corresponding operator. This is, once for the GT strength and twice for two-neutrino $\beta\beta$ decay matrix element.

\begin{figure}[b]
\begin{center}
\includegraphics[width=.82\textwidth]{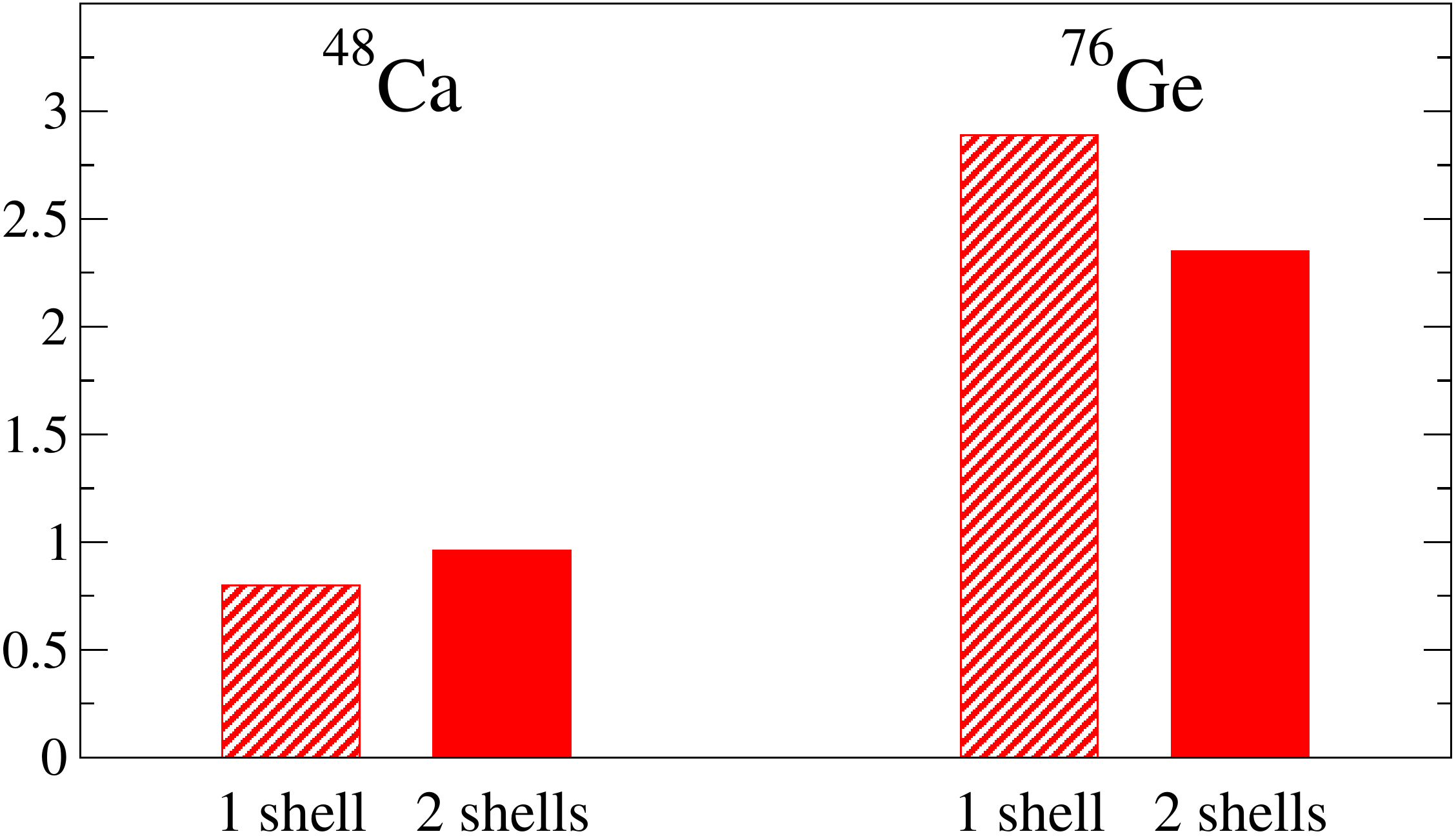}
\end{center}	
\caption{\label{fig:ca48_nme}
$^{48}\textrm{Ca}$ and $^{76}$Ge $0\nu\beta\beta$ decay
NME calculated in a one-shell (shaded) and two-shell (solid) configuration space. One-shell~\cite{menendez09} and $^{48}\textrm{Ca}$~\cite{iwata16} calculations are performed with the full shell model, while the $^{76}$Ge two shell calculation~\cite{jiao17} use the approximate generator-coordinate method.}
\end{figure}

The result of Ref.~\cite{iwata16} is shown in Fig.~\ref{fig:ca48_nme}. Note that the NME does not include a possible ``renormalization" of the NMEs, even though such ``renormalization" is required to reproduce the $^{48}$Ca two-neutrino $\beta\beta$ decay half-life. The main conclusion found in Ref.~\cite{iwata16} is that, in spite of performing the calculation in a significantly larger configuration space, the $^{48}$Ca NMEs was enhanced by only $\sim30\%$ in the two-shell calculation compared to the one-shell one. The main reason for such a relatively small effect is the competition between two type of contributions: on the one hand, pair-type two-particle--two-hole excitations into the additional harmonic oscillator shell tend to enhance the value of the NMEs~\cite{caurier08}; on the other hand, one-particle--one-hole type excitations---which are generally related to two decaying nucleons coupled to angular momentum $J>0$---tend to reduce the value of the NME~\cite{caurier08}. Overall, the competition between these two kinds of contributions results in a moderate enhancement of the NME in the expanded configuration space. Since the competition is expected to be general, similar effects are expected for extending the configuration space of shell model NME calculations in heavier $\beta\beta$ emitters.

This expectation is consistent with the recent result of Ref.~\cite{jiao17}, which calculated the nuclear matrix element of $^{76}$Ge in a configuration space consisting of two harmonic oscillator shells. 
This calculation is based on the generator-coordinated method, which does not include all the many-body correlations present in the shell model, because a full shell model diagonalization of the two-shell configuration space is beyond present computing capabilities. Fig.~\ref{fig:ca48_nme} shows the result of Ref.~\cite{jiao17} in comparison with a shell model calculation in one oscillator shell. Similarly to the findings in $^{48}$Ca, the impact of increasing the size of the configuration space is small, with the NME even slightly reducing its value in the larger space.

\section{$\beta\beta$ decay and double Gamow-Teller transitions}

In the absence of a $0\nu\beta\beta$ detection, theoretical calculations of the NMEs have to be tested against different nuclear structure data. First, all calculations compare their predictions to the nuclear structure of the initial and final states of the decay. In addition, an obvious observable to test calculations is the two-neutrino $\beta\beta$ decay, which shares initial and final states with $0\nu\beta\beta$ decay and has similar spin-isospin structure. However, the momentum transfers in the two $\beta\beta$ modes are very different: while in the two-neutrino case the momentum transfer is limited by the $Q$-value---a couple of MeV---in the neutrinoless case momentum is transferred via the not-to-be-emitted virtual neutrinos. A test of the relevant momentum-transfer regime---about $q\sim100$~MeV---would involve a comparison to muon capture or neutrino scattering. Unfortunately, data on these observables is limited.
GT transition strengths measured in charge-exchange experiments are also typically used to test calculations. Described by the same operator as GT $\beta$ decays, GT transitions are not limited by the $Q$-value, and can be studied to energies even past the GT GR at $E\sim10-15$~MeV.

A closer connection to $0\nu\beta\beta$ decay can be expected to come from double GT (DGT) transitions that are being looked for in double charge-exchange experiments~\cite{takaki-aris,uesaka-nppac,cappuzzello,takahisa-17}. The operator structure $0\nu\beta\beta$ decay of DGT transitions is very similar, with the corresponding matrix elements given by
\begin{eqnarray}
&M^{0\nu\beta\beta}(i\rightarrow f)&
=M_{GT}^{0\nu}+\frac{M_F^{0\nu}}{g_A^2}+M_T^{0\nu}= \sum_{X=GT,F,T}
\bra{f} \sum_{a,b} H_X(r_{ab})\,S_X\, \tau^+_a \tau^+_b \ket{i}\,,\\
\label{eq:2GT}
&M^{DGT}(i\rightarrow f)&
=\bra{f} \sum_{a,b}
[{\bm \sigma}_a \tau^+_a \times {\bm \sigma}_b \tau^+_b]^{\lambda} \ket{i}\,,
\end{eqnarray}
where $\bm \sigma$, $\tau$ denote spin and isospin, respectively, and $g_A$ is the axial coupling. The labels $F$ and $T$ stand for the subleading Fermi and tensor parts of the $0\nu\beta\beta$ NME, much smaller---less than 20\%---than the dominant GT piece associated to the spin structure $S_{GT}={\bm\sigma}_1\cdot{\bm\sigma}_2$. Therefore besides the small effect of the $F$ and $T$ terms, for DGT transitions to the ground state of the final nucleus---where the DGT operator can only couple to $\lambda=0$---the $0\nu\beta\beta$ and DGT operators only differ by the presence of the neutrino potential $H$, which depends on the internucleon distance $r_{ab}$. The form of the neutrino potentials is given in detail in Ref.~\cite{menendez_heavy}.
%
%The neutrino potentials appear in the $0\nu\beta\beta$ decay operator because the the neutrinos are not emitted. They depend on the distance between the decaying nucleons $r_{ab}$~\cite{engelmen_review}:
%\begin{eqnarray}
%\label{eq:nu_pot}
%%\fl
%H_X(r_{ab})
%= \frac{2R}{\pi} \int_0^\infty \!\!\! q^2 \, dq
%\frac{j_X(q\,r_{ab}) h_{X}(q)}
%{q\left(q+\mu\right)}\,,
%\end{eqnarray}
%with %${\bm q}$ the momentum transfer,
%$R=1.2A^{1/3}$~fm the nuclear radius,
%$j_F(q\,r_{ab})=j_{GT}(q\,r_{ab})=j_0(q\,r_{ab})$
%and $j_T(q\,r_{ab})=j_2(q\,r_{ab})$ spherical Bessel functions,
%and $\mu$ the closure energy~\cite{horoi-nonclosure-48ca}.
%The  $h_{X}(q)$ functions can be found in Ref.~\cite{menendez_heavy},
%with $h_{GT}=1$ to first order.

Reference~\cite{2GT0nbb} predicted the DGT strength of $^{48}$Ca, including the DGT GR using large-scale shell model calculations up to two oscillator shells. Interestingly, the energy of the resonance was found to be correlated---in the shell model calculation---to the value of the $0\nu\beta\beta$ decay NME. This relation is due to the dependence of the two observables to particle-like pairing correlations. As a consequence, a measurement of the DGT GR in $^{48}$Ca could provide an indication of the value of the NME of the same nucleus.

\begin{figure}[t]
	\begin{center}
		\includegraphics[width=\textwidth]{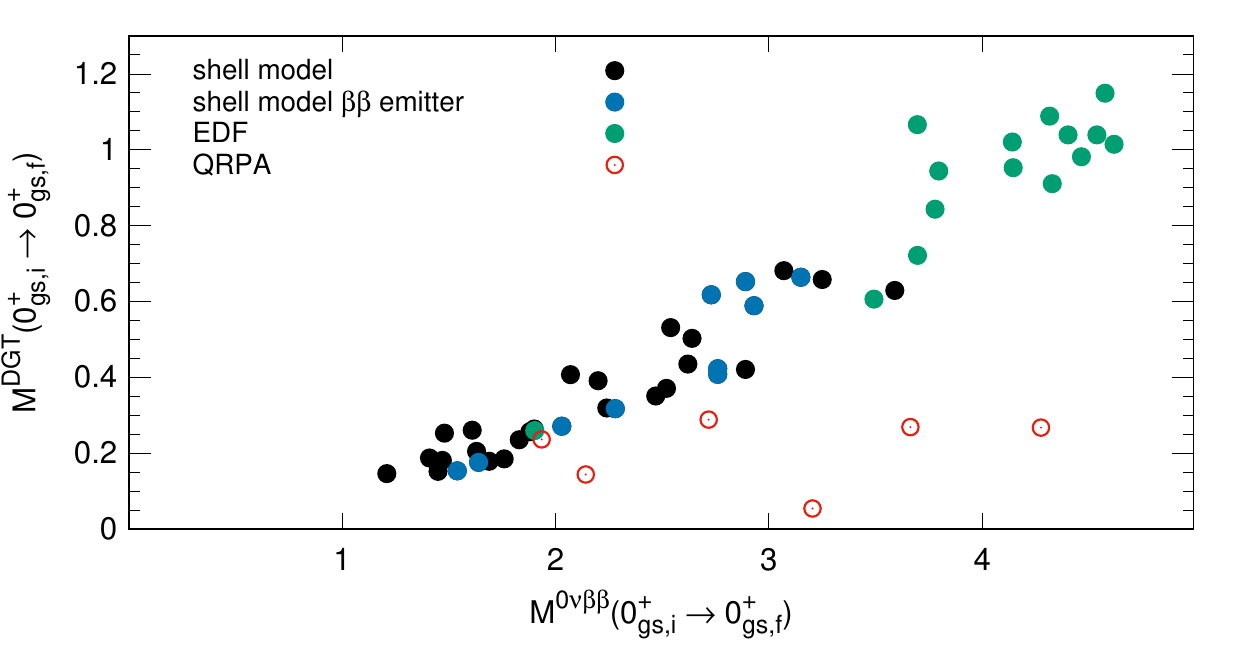}	
	\end{center}	
	\caption{\label{fig:nme_linear}	
		Correlation between
		$0\nu\beta\beta$ decay NMEs
		$M^{0\nu\beta\beta}(0^+_{gs,i} \rightarrow 0^+_{gs,f})$
		and the DGT matrix elements
		$M^{\rm DGT}(0^+_{gs,i} \rightarrow 0^+_{gs,f})$.
		Shell model results for germanium, tellurium, tin, tellurium, and xenon isotopes (black) including the $\beta\beta$ emitters $^{76}$Ge, $^{82}$Se, $^{124}$Sn, $^{130}$Te and $^{136}$Xe (blue) are compared to EDF theory~\cite{rodriguez} (green) and QRPA predictions~\cite{simkovic-11} (open red symbols).
		The calculations use several shell model interactions for each isotope~\cite{menendez09,JUN45,qi-12}.  Adapted from Ref.~\cite{2GT0nbb}.
	}
\end{figure}

In addition, Ref.~\cite{2GT0nbb} studied DGT transitions to the ground state, and compared the results to the $0\nu\beta\beta$ decay NMEs. Note that the initial and final states of both processes are the same, and also the transition operator is very similar, as discussed above. Instead of limiting to one particular case---as in the study of the DGT GR---the calculations included a set of nuclei ranging from calcium to xenon isotopes, with nuclear mass number $42\leq A\leq 136$. Therefore several $\beta\beta$ emitters but also many isotopes not relevant for $0\nu\beta\beta$ decay searches were studied. Nonetheless, these additional calculations are very useful to illuminate systematic effects.

Figure~\ref{fig:nme_linear} summarizes the results of Ref.~\cite{2GT0nbb}. A good linear correlation is found between the DGT transitions to the ground state and $0\nu\beta\beta$ NMEs. The linear correlation does not depend on the details of the shell-model interaction used, or in the correlations included in the shell-model initial and final states---as long as particle-like pairing correlations are present.  Furthermore, the correlation between $0\nu\beta\beta$ and DGT matrix elements is valid for $\beta\beta$ emitters, shown in blue in Fig.~\ref{fig:nme_linear}, and for all the other nuclei---seventeen isotopes in total. Moreover, the correlation observed in the shell model is consistent with the results of energy-density functional (EDF) theory~\cite{rodriguez}---also shown in Fig.~\ref{fig:nme_linear}---even if for the latter approach $0\nu\beta\beta$ and DGT matrix elements are much larger than the shell model ones. In contrast, quasiparticle random-phase approximation (QRPA) results~\cite{simkovic-11}---shown in Fig.~\ref{fig:nme_linear} as well---do not support the linear correlation found for the shell model.

\begin{figure}[t]
	\begin{center}
		\includegraphics[width=.95\textwidth]{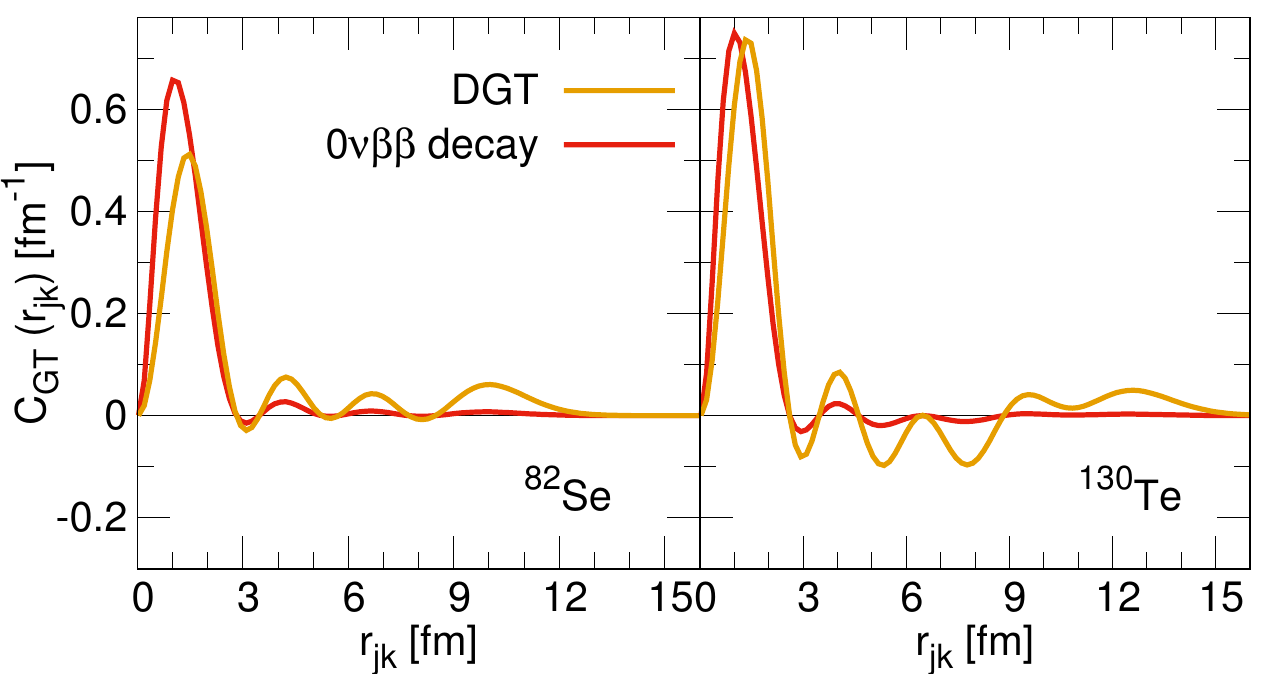}	
	\end{center}	
	\caption{\label{fig:radial_density}	
$^{82}$Se (left panel) and $^{130}$Te (right) normalized radial density distributions $C(r)$ of the GT $0\nu\beta\beta$ (red) and DGT (orange) matrix elements. Shell model interactions from Ref.~\cite{menendez09} are used.
	}
\end{figure}

The linear correlation shown in Fig.~\ref{fig:nme_linear} relates the $0\nu\beta\beta$ decay NME, driven by the weak interaction, and the DGT matrix element, a result of the strong interaction. It therefore opens the door to exploring $0\nu\beta\beta$ decay NMEs in nuclear double charge-exchange experiments~\cite{takaki-aris,uesaka-nppac,cappuzzello,takahisa-17}. This is, however, a formidable challenge at the experimental and theoretical level. First, the DGT transition is a tiny---0.03 per mil---piece of the DGT sum rule. In addition, dedicated reaction theory efforts are needed to establish the relation between double charge-exchange cross-sections and DGT matrix elements.

What is the origin of the linear correlation between $0\nu\beta\beta$ decay and DGT transitions? To address this question, Fig.~\ref{fig:radial_density} shows the normalized radial densities of the $0\nu\beta\beta$ and DGT matrix elements, defined as
\begin{eqnarray}
C_{GT}^{0\nu}(r)= \langle f | \sum_{ab} \delta(r-r_{ab}) \,H_{GT}(r_{ab})\,
{\bm \sigma}_a\cdot{\bm \sigma}_b \,\tau_a \tau_b | i \rangle / M_{GT}^{0\nu}\,, \\
C^{DGT}(r)= \langle f | \sum_{ab} \delta(r-r_{ab}) \,
[{\bm \sigma}_a\times{\bm \sigma}_b]^0 \,\tau_a \tau_b | i \rangle / M^{DGT}\,.
\label{eq:density_r}
\end{eqnarray}
Figure~\ref{fig:radial_density} shows that the two matrix elements are dominated by the contribution of nucleons that are relatively close to each other, $r_{ab}\lesssim3$~fm. In the case of DGT transitions this is the result of the partial cancellation of the longer-range contributions. This short-range dominance is non trivial, as Fig.~\ref{fig:radial_density} shows that the shell model calculation naturally probes internucleon distances up to twice the nuclear radius.

The short-range character provides an explanation for the existence of the linear correlation between the two matrix elements. The work of Bogner et al.~\cite{bogner-10,bogner-12} shows that when an operator probes only the short-range physics of low-energy states, the corresponding matrix elements factorize into a universal operator-dependent constant times a state-dependent number which is common to all short-range operators. Since both $0\nu\beta\beta$ decay and DGT shell-model matrix elements fulfill these conditions, a linear relation between them is predicted. In contrast, the QRPA DGT matrix elements receive contributions from longer range, so that the correlation is not predicted in their case, in agreement with Fig.~\ref{fig:nme_linear}.

\section{Conclusions}
We have summarized two advances that improve our understanding of $0\nu\beta\beta$ decay. On the one hand, shell model calculations in a configuration space comprising two oscillator shells suggest that the NME obtained in standard shell model calculations are reasonably converged. On the other hand, the finding of a good linear correlation between the NMEs and DGT transitions, valid across the nuclear chart, brings the opportunity to obtain precious information on $0\nu\beta\beta$ decay in double charge-exchange nuclear reactions. These advances pave the way towards a more reliable determination of the $0\nu\beta\beta$ NMEs in the mid-term future.

\section*{Acknowledgments}
I would especially like to thank Prof. T. Otsuka for many stimulating discussions and for his support, as well as for his kind introduction to research in Tokyo and Japanese culture. I am grateful to my co-authors T. Abe, M. Honma, Y. Iwata, T. Otsuka, N. Shimizu, Y.~Utsuno, and K. Yako for using in these proceedings results of our common research.
This work was supported by the
CNS-RIKEN joint project for large-scale nuclear structure calculations, and by MEXT and JICFuS as a priority issue 
(Elucidation of the fundamental laws and evolution of the universe, hp170230) 
to be tackled by using Post K Computer.


\begin{thebibliography}{9}
	
	\bibitem{KamLAND-Zen16}
	A. Gando {\em et~al.\/} (KamLAND-Zen Collaboration), {\em Phys. Rev.
		Lett.\/} {\bf 117}, 082503 (2016).
	
	\bibitem{EXO18}
	J. B. Albert {\em et~al.\/} (EXO Collaboration), {\em Phys. Rev.
		Lett.\/} {\bf 120}, 072701 (2018).
		
	\bibitem{CUORE18}
	C. Alduino {\em et~al.\/} (CUORE Collaboration), {\em Phys. Rev. Lett.\/} {\bf 120}, 132501 (2018).
	
	\bibitem{MAJORANA18}
	C. E. Aalseth {\em et~al.\/} (MAJORANA Collaboration), {\em Phys. Rev. Lett.\/} {\bf 120}, 132502 (2018).

	\bibitem{GERDA18}
	M. Agostini {\em et~al.\/} (GERDA Collaboration), {\em Phys. Rev. Lett.\/} {\bf 120}, 132503 (2018).

	\bibitem{NEXT18}
	A. D. McDonald {\em et~al.\/} (NEXT Collaboration), {\em Phys. Rev. Lett.\/} {\bf 120}, 132504 (2018).

	\bibitem{CUPID18}
	C. Azzolini {\em et~al.\/} (CUPID-0 Collaboration), arXiv:1802.07791.
	%{\em Phys. Rev. Lett.\/} {\bf 120}... (2018).

		
	\bibitem{engelmen_review}
	J. Engel and J. Men{\'e}ndez, {\em Rep. Prog. Phys.} {\bf 45}, 014003 (2017).
	
	\bibitem{cau05}
	E. Caurier, G. Mart{\'i}nez-Pinedo, F. Nowacki, A. Poves and A. P. Zuker {\em Rev. Mod. Phys.\/} {\bf 77}, 427 (2005).

	\bibitem{vogel12}
	P. Vogel, {\em J Phys. G: Nucl. Part. Phys.\/} {\bf 39}, 124002 (2012).

	\bibitem{iwata16}
	Y. Iwata, N. Shimizu, T. Otsuka, Y. Utsuno, 
	J. Men\'endez, M. Honma, and T. Abe, 
	{\em Phys. Rev. Lett.} \textbf{116}, 112502 (2016).
	
	\bibitem{iwata15}
	Y. Iwata, N. Shimizu, Y. Utsuno, M. Honma, T. Abe, and T. Otsuka, 
	{\em JPS Conf. Proc.} \textbf{6}, 030057 (2015).	

	\bibitem{yako-ca48}
	K. Yako {\it et al.}, % Phys. Rev. Lett.
	{\em Phys. Rev. Lett.} \textbf{103}, 012503 (2009).

	\bibitem{menendez09}
	J. Men\'endez,  A. Poves, E. Caurier, and F. Nowacki,
	{\em Nucl. Phys. A} \textbf{818}, 139 (2009).

	\bibitem{jiao17}
	C. F. Jiao, J. Engel, and J. D. Holt,
	{\em Phys. Rev. C} \textbf{96}, 054310 (2017).

	\bibitem{caurier08}
	E. Caurier, J. Men\'endez, F. Nowacki, and A. Poves, 
	{\em Phys. Rev. Lett.} \textbf{100}, 052503 (2008).     

%	\bibitem{barabash-15} 
%	A. S. Barabash, 
%	{\em Nucl. Phys. A} \textbf{935}, 52 (2015).

%\bibitem{zinner06}
%N.~T.~Zinner, K.~Langanke and P.~Vogel,
%Phys.\ Rev.\ C {\bf 74}  024326 (2006).
%
%\bibitem{hayes03} 
%A.~C.~Hayes, P.~Navratil and J.~P.~Vary,
%Phys.\ Rev.\ Lett.\  {\bf 91}, 012502 (2003).
%
%\bibitem{suzuki06} 
%T.~Suzuki, S.~Chiba, T.~Yoshida, T.~Kajino and T.~Otsuka,
%Phys.\ Rev.\ C {\bf 74}, 034307 (2006).
%
%	\bibitem{ichimura-06} 
%	M. Ichimura, H. Sakai, and T. Wakasa, 
%	{\em Prog. Part. Nucl. Phys.} \textbf{56}, 446 (2006).
%	
%	\bibitem{frekers-13} 
%	D. Frekers, P. Puppe, J. H. Thies, and H. Ejiri, 
%	{\em Nucl. Phys. A} \textbf{916}, 219 (2013).

	\bibitem{cappuzzello} F. Cappuzzello, M. Cavallaro, C. Agodi, 
	M. Bondi, D. Carbone, A. Cunsolo and A. Foti,
	{\em Eur. Phys. J. A} \textbf{51}, 145 (2015).
	
	\bibitem{takaki-aris} % 12C(18O,18Ne)12Be
	% http://journals.jps.jp/doi/abs/10.7566/JPSCP.6.020038
	M. Takaki {\it et al.}, {\em JPS Conf. Proc.} \textbf{6}, 020038 (2015).
	
	\bibitem{uesaka-nppac}
	T. Uesaka {\it et al.}, RIKEN RIBF NP-PAC, NP1512-RIBF141 (2015).
	
	\bibitem{takahisa-17}
	K. Takahisa {\it et al.}, arXiv:1703.08264. 

%	\bibitem{horoi-nonclosure-48ca}
%	R. A. Sen'kov and M. Horoi, {\em Phys. Rev. C} \textbf{88}, 064312 (2013).
%
	\bibitem{menendez_heavy}
	J. Men{\'e}ndez, {\em J. Phys. G. Nucl. Part. Phys.} {\bf 45}, 014003 (2018).

	\bibitem{2GT0nbb}
	N. Shimizu, J. Men{\'e}ndez, and K. Yako, %arXiv:1709:01088.
	{\em Phys. Rev. Lett.\/} {\bf 120}, 142502 (2018).

	\bibitem{rodriguez} 
	T. R. Rodr\'iguez, G. Mart\'inez-Pinedo, {\em Phys. Lett. B} \textbf{719}, 
	174 (2013).
	
	\bibitem{simkovic-11}
	F. {\v S}imkovic, R. Hod\'ak, A. Faessler, and P. Vogel, 
	{\em Phys. Rev. C} \textbf{83}, 015502 (2011).
	

    \bibitem{JUN45}
    M. Honma, T. Otsuka, T. Mizusaki, and M. Hjorth-Jensen, {\em Phys. Rev. C} \textbf{80}, 064323 (2009). 
    
    \bibitem{qi-12}
    C. Qi, and Z. X. Xu, {\em Phys. Rev. C} \textbf{86}, 044323 (2012).

	\bibitem{bogner-10}
	E. R. Anderson, S. K. Bogner, R. J. Furnstahl, and R. J. Perry, 
	{\em Phys. Rev. C} \textbf{82}, 054001 (2010).
	
	\bibitem{bogner-12}
	S. K. Bogner, and D. Roscher,
	{\em Phys. Rev. C} \textbf{86}, 064304 (2012).    
				
	
%	\bibitem{Pastore17b}
%	S. Pastore, J. Carlson, V. Cirigliano, W. Dekens, E. Mereghetti and R.~B. Wiringa, arXiv:1710.05026.
	
	
	
\end{thebibliography}
\end{document}